# Computational Kerr-Ellipsometry: Quantifying Broadband Optical Nonreciprocity of Magneto-Optic Materials

Vishal Choudhury[1], Chinmay Khandekar[2], Ashwin K. Boddeti[2], Ali Jishi[2], Mustafa Erkovan[3], Tyler Sentz[2], Farid Kalhor[2], Susana Cardoso[4,5], V. R. Supradeepa[6, *], Zubin Jacob[2, *]

[1] *Max Planck Institute for the Science of Light, Staudtstraße 2, 91058 Erlangen, Germany*

[2] *Birck Nanotechnology Center, Elmore Family School of Electrical and Computer Engineering, Purdue University, West Lafayette, Indiana 47907, USA*

[3] *Department of Fundamental Sciences and Engineering, Sivas University of Science and Technology, Sivas, Turkey*

[4] *INESC Microsistemas e Nanotecnologias, Rua Alves Redol, 9, 1000-029 Lisboa, Portugal*

[5] *Instituto Superior Tecnico, Universidade de Lisboa, Av. Rovisco Pais, 1000 Lisboa, Portugal*

[6] *Centre for Nano Science and Engineering, Indian Institute of Science, Bengaluru 560012, India*

*Corresponding author. zjacob@purdue.edu

*Corresponding author. supradeepa@iisc.ac.in

**Abstract**

Characterizing the optical response of magneto-optic and magnetic materials usually relies on semi-classical models (e.g. Lorentz oscillator model) involving few parameters or models based on a detailed quantum mechanical description of the underlying response. These models typically involve a few parameters that are estimated via fitting the experimental data to provide a qualitative understanding of the underlying physics. Such a few-parameters fitting approach falls short of accurately capturing all elements of the complex-valued permittivity tensor across a range of wavelengths. Accurate characterization of the permittivity tensor elements across a broad range of wavelengths is invariably imperative for designing optical elements such as isolators, circulators, etc. Here, we propose and demonstrate a ubiquitous and accessible method based on a combination of spectroscopic ellipsometry and spectroscopic Magneto-Optic Kerr Effect (MOKE) measurements coupled with rigorous numerical parameter extraction techniques. To this end, we use the combined MOKE ellipsometry measurements conducted at different angles of incidence with a gradient-descent minimization algorithm to provide the inverse solution to the complete dielectric permittivity tensor. Further, we demonstrate model re-verification to ensure the estimated dielectric permittivity values reliably predict the measured experimental data. Our method is a simplified bench-top counterpart to the otherwise complex measurement systems.

## I. INTRODUCTION

Over the last several decades, magneto-optic (MO) materials have revolutionized data storage devices and other optical technologies. Magnetic anisotropy in MO materials stems from strong Spin-Orbit coupling between the constituent elements within the crystal structure [1]. These materials play an important role in the lasers and telecommunications industry, primarily being used in optical isolators and circulators [2]. Similarly, reciprocal MO materials exhibit optical activity and are also of interest in spectroscopy and the sugar industry [3, 4]. MO materials are also widely investigated in the fields of spintronics [5], magneto-plasmonics [6] and light-matter interaction [7]. Heat-assisted magnetic recording (HAMR) and all-optical switching (AOS) [8, 9] in spintronic materials are currently sought-after pathways to further advance data storage technology. More recently, topological metamaterials exhibiting optical nonreciprocity have been of substantial interest [10]. Topological activity in materials operating in visible and near-infrared (IR) has been reported by multiple groups [11, 12, 13]. Similarly, topological Dirac and Weyl semimetals exhibiting giant MO activity have been of interest [14, 15]. In the linear regime, the MO properties of all the aforementioned materials can be characterized by the linear dielectric tensor. Thus, it is of utmost importance to have techniques that can provide ease and accuracy in the measurement of the dielectric tensor of materials spanning all the domains.

Evaluating the diagonal elements of the dielectric tensor using spectroscopic ellipsometry is a well-established technique. Currently, many commercial ellipsometers are widely available for operations over hundreds of nanometer bandwidth between ultraviolet (UV) to IR [16]. Similarly, commercial ellipsometers in their generalized ellipsometry (GE) mode can be used to characterize the off-diagonal elements of the dielectric tensor [17, 18]. On the other hand, for samples exhibiting depolarizing or birefringent effects, often using Stokes vector formalism to calculate the Mueller matrix can be useful [18]. Such GE systems can be further upgraded for field dependent as well as magnetic anisotropy determination measurements by integrating an external magnetic field [19, 20]. Additionally, vector magneto-optical generalized ellipsometer (VMOGE) [21] has been demonstrated where the magnetic field can be applied in multiple orientations. This was enabled by the addition of an octupole magnet into the GE setup.

The off-diagonal parameter extraction techniques implemented in all the aforementioned methods rely on applying an external magnetic field along each of the directions for the respective parameter. Naturally that requires using rotation mounts for each direction in the sample and the magnet or using an octupole magnet as in the case of VMOGE. Thus the system complexity and cost have only scaled up to meet with the demand for obtaining optical nonreciprocity parameter $g$, as a function of different parameters and orientations. Because of additional system complexity and cost, these approaches are not readily available to researchers and there is a need for developing simpler and cheaper alternatives.

The most important bottleneck however comes in the extraction of the dielectric coefficients from the experimental data. Wavelength-dependent semi-classical models with few parameters can be used to characterize the magneto-optic response of Drude-Lorentz materials like doped semiconductors in a magnetic field. But the same models cannot be used for all classes of magneto-optic and magnetic materials as they often require detailed quantum descriptions of the underlying response. The few-parameters models are insufficient to accurately capture all complex-valued permittivity tensor elements at all wavelengths. Therefore, optimization algorithms are required at every single wavelength to extract the underlying magneto-optic parameters that provide the best fit to the experimentally measured data. However, even this approach requires careful consideration of the multi-valuedness of the optimized solution. The same algorithm can converge to multiple distinct solutions depending on different initial guess values, thereby significantly reducing the confidence in the obtained solution.

In this work, we present a new method to probe the dielectric tensor of all forms of MO materials using Magneto Optic Kerr Effect (MOKE) as well as ellipsometry measurements. Using a UV-vis-NIR spanning laser and an in-house built electromagnet, we conduct MOKE measurements on GdFeCo samples with in-plane magnetic anisotropy (IMA) and extract the entire dielectric tensor. GdFeCo is of keen interest in ultrafast AOS due to its ability to switch with a single pulse excitation [22, 23]. The AOS property already demonstrated across visible, near and mid-IR [23, 24], makes the characterization of optical parameters even more relevant for such materials. AOS behavior, CMOS compatibility as well as its low saturation magnetization makes GdFeCo the most relevant MO material to propose our new method. MOKE setups are widely available across the community and therefore, this technique can be readily used with existing MOKE setups for material characterization. The model dielectric tensor for the GdFeCo layer is obtained when the minimum value in a predefined merit function is obtained. We further verify the extracted gyrotropy parameters by predicting the MOKE values that are not used in the extraction algorithm. This step is akin to model verification in multi-target regression and machine learning methods. This is in contrast to some of the recent works on magneto-optic ellipsometry [25] where MOKE characterization in the Fourier domain was performed to extract the magnetization vector of a sample. In our work, the calculation of Fresnel coefficients from Maxwell's equations is exact, while some form of linear approximation is used in [25]. In another recent work [26], the accuracy of Fresnel coefficients in linear magneto-optics was experimentally investigated. However exact calculation of Fresnel coefficients enables us to extract the dielectric tensor itself, which is a more fundamental parameter. This also makes it possible to calculate the Fresnel coefficient at arbitrary orientations thereby accurately predicting arbitrary MOKE values as well. To the best of our knowledge, model verification via prediction of experimental data not used for fitting has not been demonstrated in previous magneto-optic characterization works which are limited to the least-squares fitting of the obtained data. It is important to add that the 'linearity' in the dielectric tensor may not always be true and for emerging magneto-optic materials exhibiting novel effects, other 'nonlinear' (e.g. cubic) dependencies could potentially be expected. We believe, exactly solving Fresnel's coefficients will enable the understanding of this deeper connection between the magnetization and the optical response. As a result, high-fidelity dielectric response characterization techniques are crucial and that is the main focus of this work. Thus, in the current scenario, though the sample used in our experiments has an IMA, this method is general enough to be extended for samples with perpendicular as well as cubic magnetic anisotropy. In the end, we also explore the possible effects of nonlinearity using azimuthally resolved characterization measurements, the results of which are also discussed at the end of this work.

## II. EXPERIMENT AND NUMERICAL METHOD

### A. Experiment

The experiments were conducted in two parts; first, the sample was mounted in the in-house built spectroscopic MOKE setup (Fig. 1 (b)) which implements lock-in detection in a balanced detection configuration. The lock-in detection at a high modulation frequency on the signal from the balanced detector enabled a noise floor of ~100 μrad. Longitudinal MOKE measurements were carried out for multiple angles of incidence as well as at multiple azimuthal orientations. A maximum magnetic field of 50 Gauss was applied in the measurements, as the VSM measurements showed a coercive field of less than 15 Gauss for the easy axis (supplementary material). Additionally, the configurations of the setup to conform with multiple angles of incidences limited the field applied to 50 Gauss along the plane of the sample for longitudinal configuration. A broadband probe laser delivering ~8 nm linewidth of individual probe wavelength enabled spectroscopic MOKE to be carried out between 400 nm to 800 nm. A raspberry-pi controller interface enabled the recording of MOKE measurements for a user-defined number of sweeps of the magnetic field at any desired azimuthal or polar angles of incidences. Further details of the experimental setup are described in the supplementary material. In the next step, standard ellipsometer measurements were also conducted on the sample at different angles of incidence. The sample was prepared by depositing a 20 nm $(Fe_{10}Co_{90})_{0.74}Gd_{0.26}$ thin film on Si/SiO2 substrate using magnetron sputtering (supplementary material).

### B. Numerical method

The linear dielectric tensor for MO materials is given as

$$\epsilon + g = \begin{bmatrix} \varepsilon_x & 0 & 0 \\ 0 & \varepsilon_y & 0 \\ 0 & 0 & \varepsilon_z \end{bmatrix} + \begin{bmatrix} 0 & ig_z & -ig_y \\ -ig_z & 0 & ig_x \\ ig_y & -ig_x & 0 \end{bmatrix} \qquad (1)$$

where, $\epsilon$ and $g$ are the diagonal dielectric tensor and the dielectric tensor containing the off-diagonal elements as denoted by the first and second matrices respectively. Here, $g_i = Q_i M_i$ ($i = x, y, z$) is called linear optical gyrotropy coefficient where $\vec{M} = M_x + M_y + M_z$ and $\vec{Q} = Q_x + Q_y + Q_z$ are the magnetization in the material and optical coupling constants respectively [21].

The two different experimental measurements of ellipsometry and MOKE depend on the Fresnel reflection coefficients $r_{jk}$ (where $j$, $k$ stand for $s$ and $p$ – polarization) denoting the amplitude of $j$ – polarized reflected light due to unit-amplitude incident $k$ – polarized light of given wavelength and at given angles of incidence. In particular, the ellipsometry data ($\Psi, \Delta$) depends on the reflection coefficients as:

$$\Psi = \tan^{-1}\left|\frac{r_{pp}}{r_{ss}}\right|; \quad \Delta = angle\left(\left|\frac{r_{pp}}{r_{ss}}\right|\right) \quad (2)$$

And, the MOKE data includes Kerr rotation and Kerr ellipticity which are defined as the following

$$\theta_k = real\left(\tan^{-1}\left(\frac{r_{sp}}{r_{pp}}\right)\right); \quad \eta_k = imag\left(\tan^{-1}\left(\frac{r_{sp}}{r_{pp}}\right)\right) \quad (3)$$

To predict the ellipsometry and MOKE measurements, the required Fresnel coefficients dependent on the underlying permittivity tensor describing the magneto-optic material (Eq. (1)) are calculated using the well-known transfer matrix (TMM) approach, generalized to magneto-optic materials. For our method of calculation of Fresnel coefficients, we use the generalized version applicable for all types of bi-anisotropic materials that was developed in our other recent work [27]. It is employed for the prediction of the experimentally measured data and is available for everyone at [link:].

To ensure a high accuracy of prediction from the numerical model, we employ a two-step approach to the extraction of the permittivity tensor. We employ ellipsometry-informed initial guess of diagonal parameters and use sufficiently many MOKE measurements at different angles of incidences to ensure that the optimization converges to the same solution from different initial guess values of the gyrotropy parameters. We use $N' = 13$ number of MOKE measurements to predict $12 (= N)$ real-valued gyrotropy parameters (6 diagonal and 6 off-diagonal in Eq. (1)). Having lesser number of measured MOKE data $N'(< N)$ as opposed to the number of parameters to be predicted often converges the optimization to more than one solution. The numerical method for extracting $\epsilon$ and $g$ predominantly consists of the following two steps in the given order,

  I. Extract only $\epsilon$ using ellipsometry data as the fitting set.

II. Extract $g$ using MOKE data as the fitting set with the extracted $\epsilon$ as the initial guess.

In both of the above steps, two different merit functions ($MF$) are defined and the optimization algorithm continues such that both $MF$s are minimized. As depicted in Fig. 2, the ellipsometry data ($\Psi, \Delta$) acquired for the respective angles of incidences ($\theta_i$) are used to fit to the model dielectric tensor by optimizing the inverse solution via gradient descent. Using the angle of incidences and the initial guess of the model dielectric tensor, the Fresnel reflection coefficients and subsequently a set of modeled ellipsometer data are calculated using TMM for a converging $MF$. The $MF$ here is defined by

$$MF_{ell} = \sum_i^N [|(\Psi_i^{calc} - \Psi_i^{meas})| + |(\Delta_i^{calc} - \Delta_i^{meas})|] \qquad (4)$$

Where, $\Psi_i^{calc}$, $\Delta_i^{calc}$ and $\Psi_i^{meas}$, $\Delta_i^{meas}$ are the numerically calculated and experimentally measured ellipsometer data respectively. $\Psi_{calc}$ and $\Delta_{calc}$ are obtained from Eq. (2). The point at which $MF_{ell}$ converges to a minima, the model dielectric tensor is used to predict the ellipsometer data at an arbitrary angle of incidence not used in the optimization process. An accurate comparison between the predicted and measured data concludes the task of extracting the diagonal elements.

In the next step, using the measured MOKE data at different $\theta_i$ as well as the extracted $\epsilon$ from the previous step as initial guess, all the elements of $\epsilon$ and $g$ are extracted. The extracted data are then used to calculate a set of modelled MO Fresnel coefficients ($r_{sp}$ and $r_{ps}$) and subsequently the MOKE angles. $MF$ used in here is defined as

$$MF_{MOKE} = \sum_i^N |(\theta_{k_i}^{calc} - \theta_{k_i}^{meas})| \qquad (5)$$

Where $\theta_{k_i}^{calc}$ and $\theta_{k_i}^{meas}$ are the calculated and measured Kerr rotations. Similarly for this case, Kerr rotation and Kerr ellipticity is defined using Eq. (3). As depicted in Fig. 2, when $MF_{MOKE}$ converges to a minima, MOKE angles at arbitrary angle of incidences are predicted and subsequently compared with measured data for accuracy.

The constituent novel aspects in our numerical method are of ensuring convergence to the same solution with different initial guesses with sufficiently many MOKE measurements as well as an additional step of model verification using unused MOKE data. These two collectively provide more confidence in the extracted gyrotropy parameters compared to the previously employed least-squares fitting. The same procedure is repeated for entire sweeps of MOKE data acquired at different wavelength of light and azimuthal orientation of the sample. Thus, the entire dielectric tensor of the material can be attained as a function of wavelength, magnetic field and azimuthal angle.

### III. RESULTS AND DISCUSSION

Due to the lack of birefringence in the sample, only two diagonal elements ($\varepsilon_x^{real}, \varepsilon_x^{imag}$) of the dielectric tensor is extracted. Figure 3 (a) shows the extracted real and imaginary parts of the extracted diagonal element between 400 nm and 1000 nm. The negative value of the real part indicates a metallic nature in the thin film (20 nm). Due to isotropy, the diagonal elements of the dielectric tensor are also invariant under any azimuthal rotation in the sample. As evident from the figure, the sample does not exhibit any Lorentzian spectral features over the visible and NIR wavelength range investigated. But more importantly, observation of any possible spectral feature originating from Lorentzian or non-Lorentzian origin is also limited by the laser linewidth. This would require using pump-probe measurement setup with finer linewidth swept across the bandwidth, which is beyond the scope of this work. At an incidence angle of $\theta_i = 50^0$, we calculated the ellipsometry data ($\Psi_i^{calc}, \Delta_i^{calc}$) using the now extracted diagonal dielectric tensor using Eq. (2). The measured ellipsometer data ($\Psi_i^{meas}, \Delta_i^{meas}$) at $\theta_i = 50^0$ was not part of the set of data used in the fitting process and thus was only used in the prediction step as indicated in Fig. 2. In Fig 3 (b) and (c), a comparison between measured ($\Psi_i^{meas}, \Delta_i^{meas}$) and calculated ($\Psi_i^{calc}, \Delta_i^{calc}$) ellipsometer data for the case of $\theta_i = 50^0$ incidence shows a very high accuracy. The contents of Fig. 3 including data extraction and prediction were obtained in a few 10s of seconds owing to only two unknowns to be extracted. The computation time required for extracting the diagonal dielectric parameters for birefringent material is expected to be longer.

Once the diagonal elements are obtained, the off-diagonal data extraction is initiated. As mentioned in the previous section, $\varepsilon_x^{real}, \varepsilon_x^{imag}$ extracted in the first step are used as the initial values in the subsequent steps. In this step a total of 12 parameters, which included 6 parameters representative of $\varepsilon_i$ (3 real and 3 imaginary) and 6 $g_i$ (3 real and 3 imaginary) were optimized and extracted. Optimizing the diagonal parameters again together with the off-diagonal parameters is necessary for observing effects such as magneto-plasmonic activity [28] or magneto-refractive effect [29] where $\varepsilon_i$ is influenced by $g_i$. Although no variations in $\varepsilon_i$ was observed in our case. Similarly convergence of $MF_{MOKE}$ for as many as 12 parameters also conveys the robustness of this method. The extracted $g_i$ values are presented in Fig 4.

The data presented in Fig 4 was acquired at 4 different wavelength bands across visible and NIR as a representative of the broadband nature of our measurement. In all cases, the measured MOKE angles were above the noise floor of our setup and thus the off-diagonal parameters were extracted with high confidence. The $g_x$ and $g_y$ data presented in 4 (a), (b), (c) and (d) represents the IMA in the sample. Similarly $g_z = 0$ in 4 (e) and (f) indicates the lack of PMA in the sample. Vibrating sample magnetometer (VSM) measurements (supplementary material) conducted on the sample also shows the similar values of coercive field and consistent easy ($g_x$) and hard axis ($g_y$) orientations. ($g_z$) is observed to be zero across all wavelength for 50G of magnetic field. This is a consequence of the absence of PMA, something also evident in VSM measurement even at strong fields (supplementary material). Additionally, a change in sign of the real part of $g_x$ and $g_y$ is observed at 405 nm. Thus for a specific case of $\varphi = 0^0$, $H = 50\ G$, $\lambda = 810\ nm$, i.e. $\varepsilon(50, 810, 0^0)$

$$\epsilon + g = \begin{bmatrix} -7.86 + 21.05i & 0 & 0.29 + 0.32i \\ 0 & -7.86 + 21.05i & -0.50 - 0.46i \\ -0.29 - 0.32i & 0.50 + 0.46i & -7.86 + 21.05i \end{bmatrix}$$

We further extracted only 6 independent $g_i$ (3 real and 3 imaginary) parameters assuming the antisymmetric nature, $g^T = -g$ of the gyrotropy matrix. Additionally, using the same set of data as used in the previous steps, the calculations were repeated without assuming $g$ as antisymmetric. This time the calculations were done for 12 independent $g_i$ values (6 real and 6 imaginary). The $\varepsilon_i$ values were not part of this set of calculations as they have already been established to not change during calculations. The calculated 12 independent $g_i$ values subsequently resulted in the same antisymmetric $g$ matrix, something that is expected. For the case of extracting 6 independent $g_i$ (3 real and 3 imaginary) parameters after extracting the $\varepsilon_i$, it took less than 10 seconds to calculate one set of 6 parameters for a specific set of $\varphi$, $H$, $\lambda$ values. This computation time is relatively slower compared to the case of extracting 2 diagonal parameters owing to more unknowns to be extracted. This calculation reiterates the robustness of the TMM based numerical method where we observe the convergence of multiple parameters. Thus, this method can also be extended to samples with PMA as well as biaxial magnetic anisotropy.

Finally, in order to verify the accuracy of the complete extracted dielectric tensor, MOKE plots for the case of angle of incidence, $\theta_i = 30^0$ were predicted and compared with the measured data. Despite not using the MOKE data for $\theta_i = 30^0$ in fitting, the predicted data stands with very high accuracy in comparison with the measured data as shown in Fig 5. It is important to add that for all orientations of the incident laser, MOKE angles of less than 1 mrad were recorded. Thus, all measurements were conducted close to the noise floor of the system. The high confidence in the data in Fig 5. indicates the reliability of using exact calculation of Fresnel coefficients from measured data. We believe for materials such as Weyl semimetals [14] exhibiting 10s of mrad Kerr rotation, our method will extract the dielectric coefficients with high confidence as well.

In earlier reports on dielectric tensor characterization, an external magnetic field would be applied in the direction of the specific off diagonal coefficient in order to access them. Thus we extended our method to observe the effect of azimuthal rotation $\varphi$ of the sample on $g$. In Fig. 6 we show how azimuthal rotation $\varphi$ influences $g_x$ for the case of $\lambda = 810\ nm$.

For azimuthal angles $0 < \varphi < \frac{\pi}{2}$, evaluating an effective $g_i(\varphi)$ expression is not quite straightforward. For example, one could intuitively use the rotation matrix about an angle $\varphi$ on $g_i$ (i.e. $g_x = Q_x M_x cos\varphi$; $g_y = Q_y M_y sin\varphi$) to analytically predict $g_i(\varphi)$ as it also conforms with the azimuthal symmetry relation $g_i(\varphi + \pi) = -g_i(\varphi)$. However, the calculations did not lead to consistent values (supplementary material). Since $g$ is a tensor, the mutually independent parameters $Q_i$ and $M_i$ will contribute their individual projections for the case of an arbitrary $0 < \varphi < \frac{\pi}{2}$. Additionally, the inclusion of nonlinearity terms needs to be investigated for further accurate formalism.

### IV. CONCLUSION

We have demonstrated a new and ubiquitous method for broadband optical characterization of magneto-optic materials. Our technique relying on both ellipsometry and MOKE measurements can be used for both non-magnetic and a large gamut of MO materials.

We utilize the wide availability and cost efficiency of a MOKE setup to tackle the shortcomings of the state of the art. Gradient descent based optimization, used in the numerical calculations in our method, is known to be multivalued. The uniqueness of the inverse solution is ensured by using sufficiently many values of experimental measurements and by using different randomly chosen initial guesses. A high confidence in our method is established by the additional step of predicting arbitrary data not used as fitting data. Finally, our method can also be extended to studying the gyrotropy as a function of azimuthal angle, which is of high importance for research on topological materials.

**ACKNOWLEDGMENTS**

We would like to thank Prof. Xianfan Xu and Dr. Vasudevan Iyer for the initial discussions in setting up the MOKE setup. This work was supported by Purdue-India Overseas Visiting Doctoral Fellowship of Science and Engineering Research Board (SERB) of India and U.S. Department of Energy, Office of Basic Energy Sciences under DE-SC0017717. INESC-MN wish to acknowledge the Fundação para a Ciência e a Tecnologia for funding of the Research Unit (UID/05367/2020) through plurianual BASE and PROGRAMATICO financing.

**Figure legends**

**FIG. 1. (a)** This work vs the state of the art. **(b)** Schematic of the experimental setup. A broadband UV-vis-NIR laser is integrated to a MOKE setup based on balanced and lock in detection.

**FIG. 2.** Steps involved in the numerical optimization.

**FIG. 3. (a)** Real and imaginary parts of the diagonal dielectric constants extracted from ellipsometer data, **(b)**, **(c)** Predicted ellipsometer parameters $\Psi, \Delta$ respectively using the data from **(a)** for $\theta_i = 50^0$.

**FIG. 4.** Real and imaginary parts of the off-diagonal dielectric parameters extracted from MOKE data. The hysteresis curves in **(a)**, **(b)** across all wavelength indicate the easy axis along the x-direction in our frame of reference and similarly **(c)**, **(d)** data indicates the hard axis to be along y axis in the sample. Zero value of $g_z$ in **(e)**, **(f)** confirms the lack of PMA in the sample.

**FIG. 5.** MOKE traces predicted using the extracted $g_i$ in Fig. 4 at $\theta_i = 30^0$. Comparison of the MOKE traces between measured and calculated **(a)** easy axis and **(b)** hard axis.

**FIG. 6.** Real **(a)** and imaginary **(b)** parts of $g_x$ for varying azimuthal angle $\varphi$. $\varphi = 0^0$ corresponds to the case of applying the magnetic field along the easy axis, while $\varphi = 90^0$ corresponds to applying it along the hard axis.

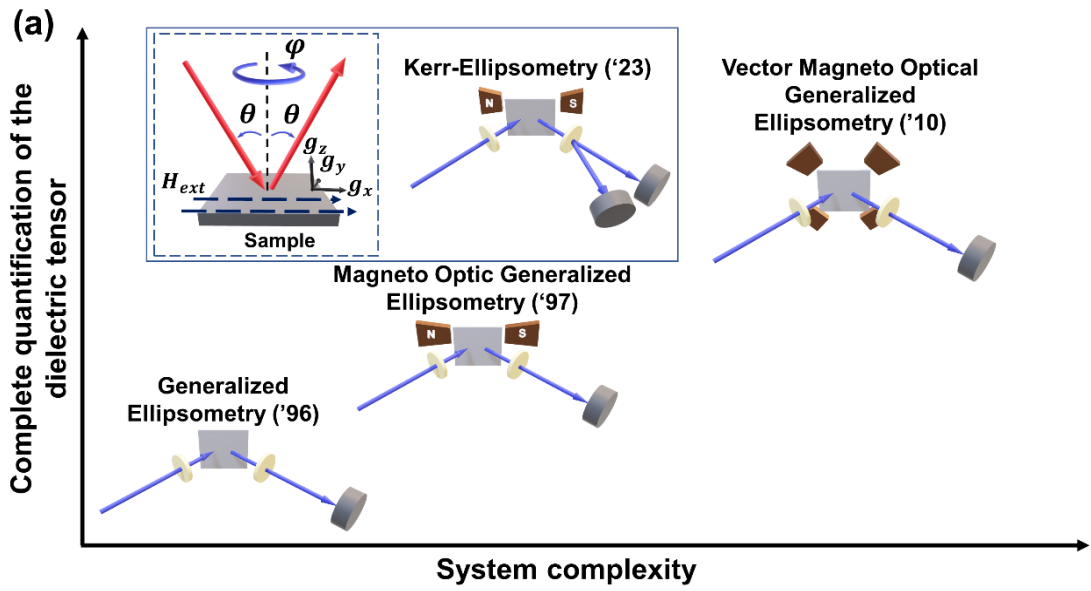

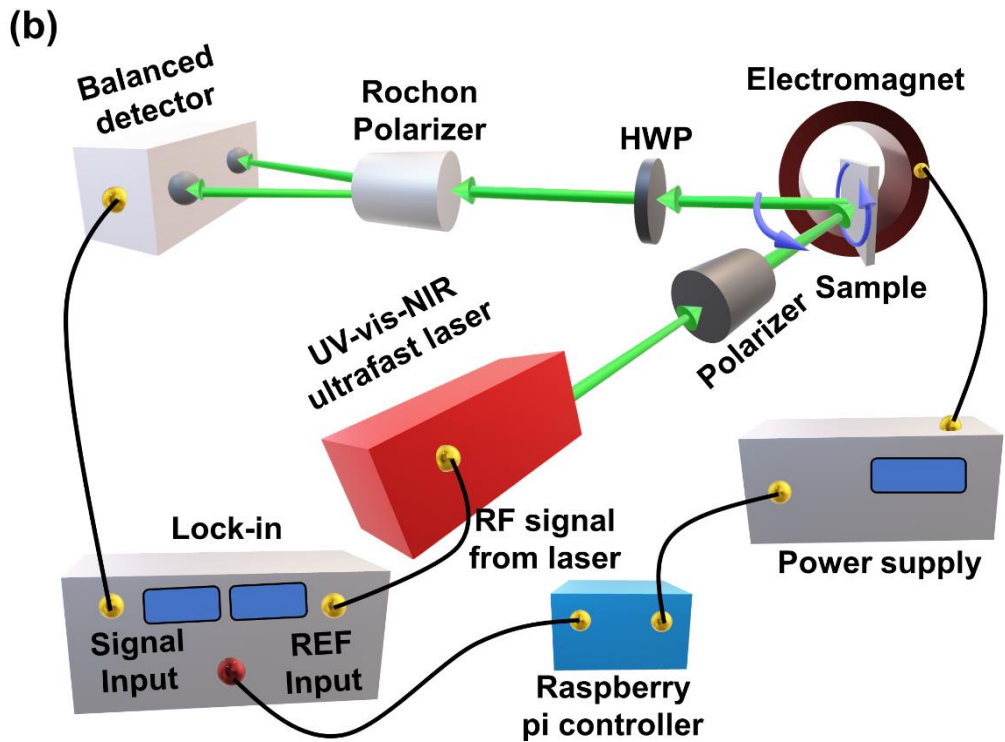

Fig. 1, V. Choudhury *et al.*

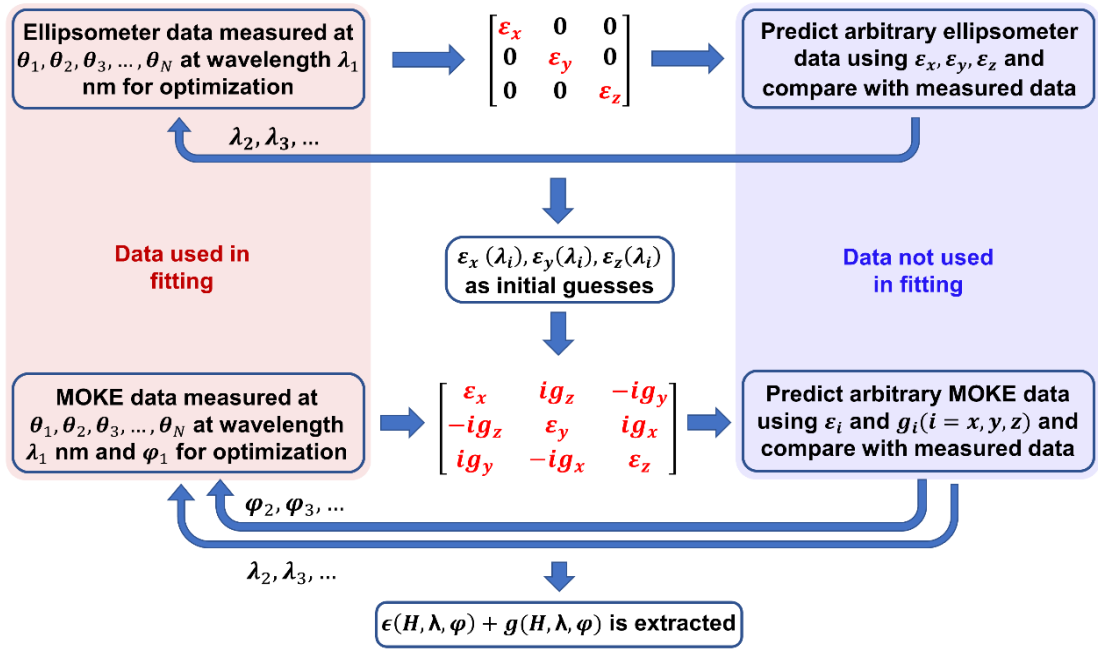

Fig. 2, V. Choudhury *et al.*

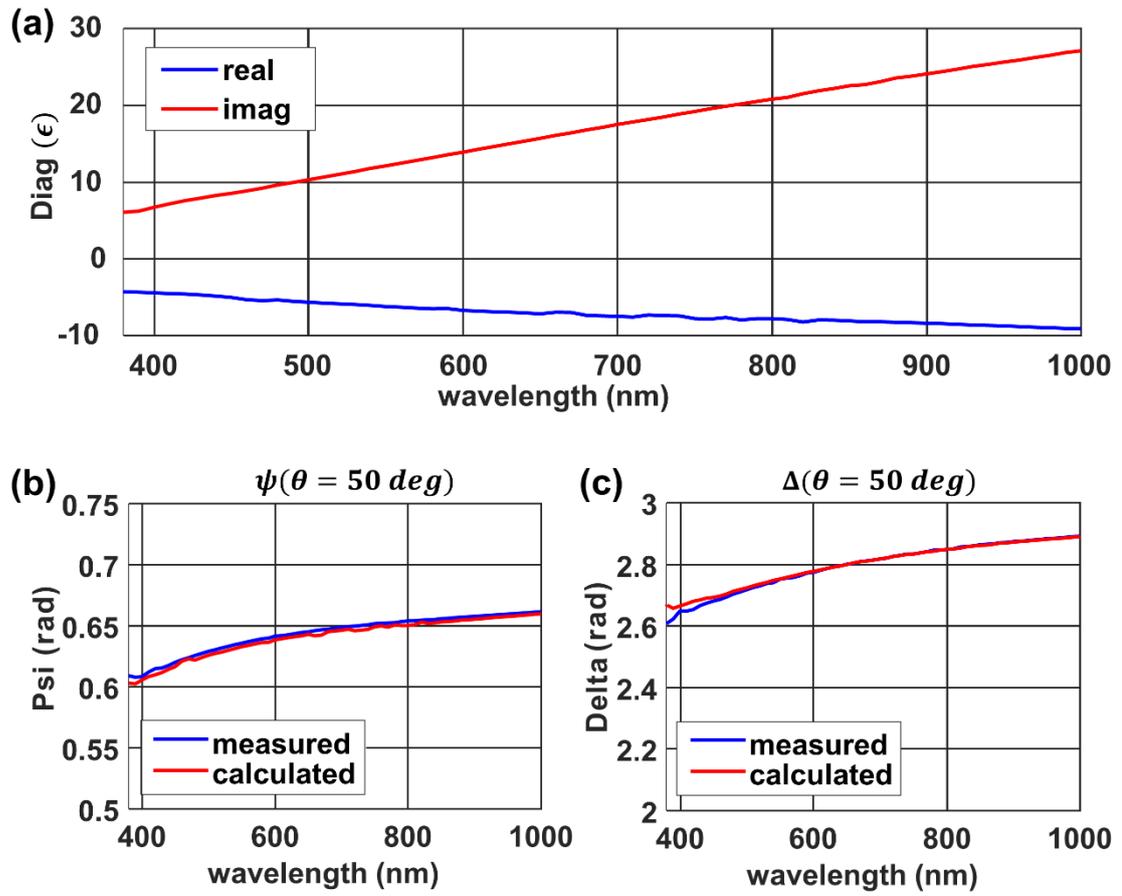

Fig. 3, V. Choudhury *et al.*

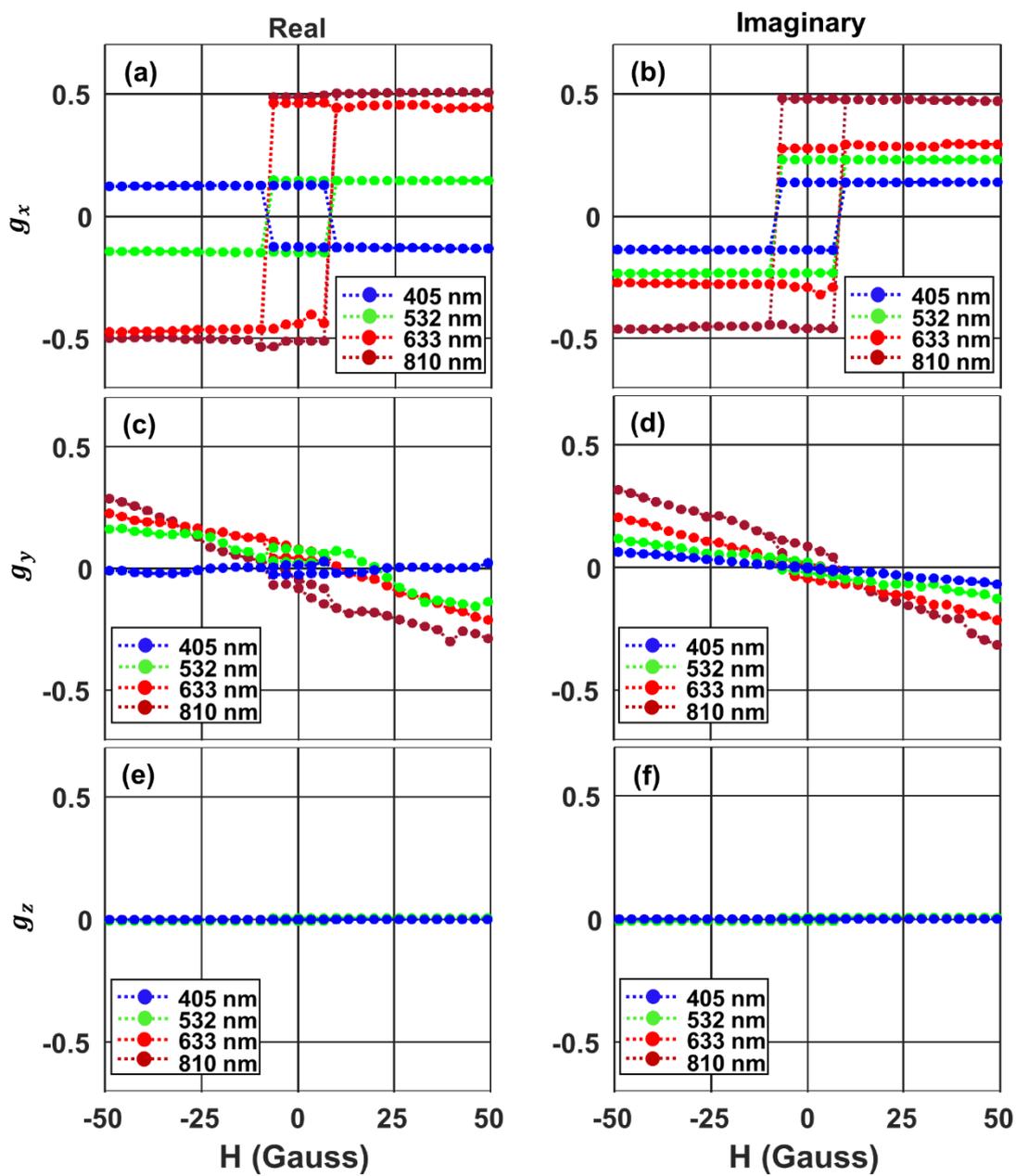

Fig. 4, V. Choudhury *et al.*

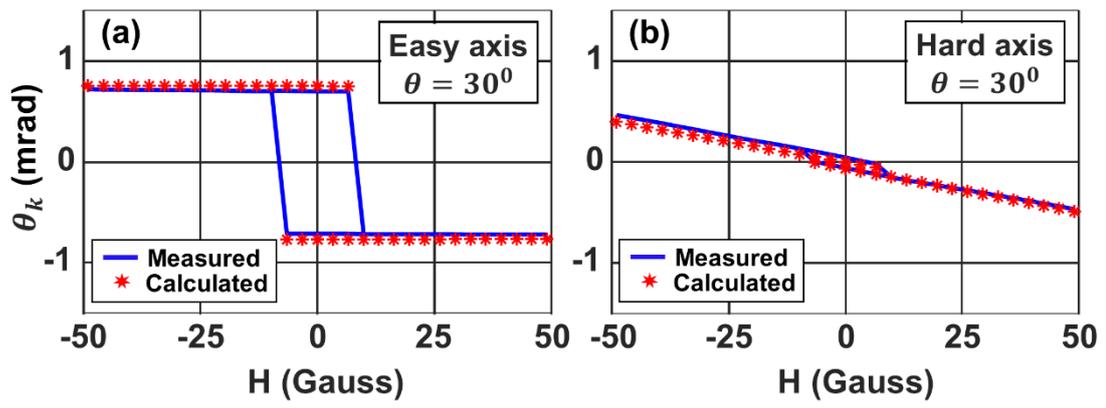

Fig. 5, V. Choudhury *et al.*

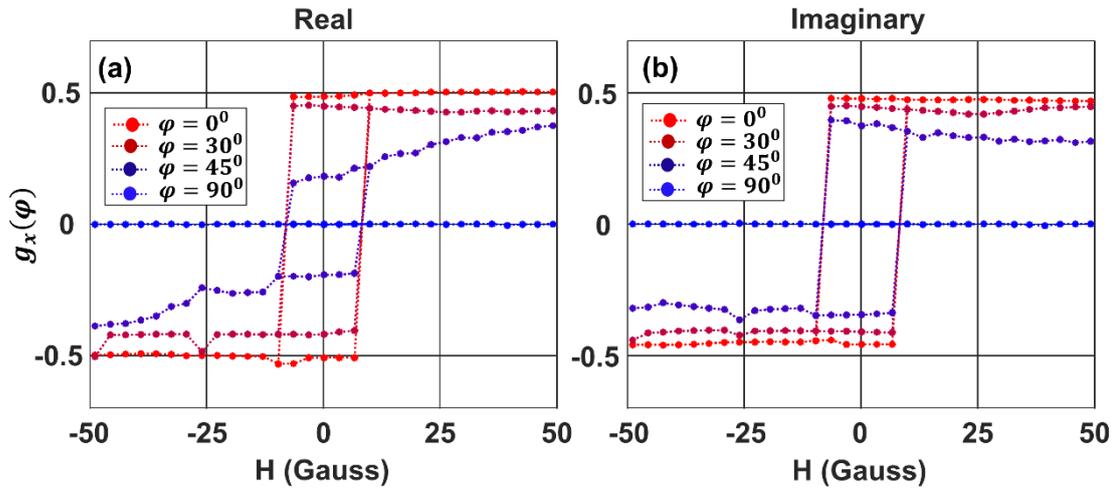

Fig. 6, V. Choudhury *et al.*